\documentclass[twoside,12pt]{article}
\usepackage{amssymb,endnotes,epsfig}
\newcounter{muni}
\newenvironment{remunerate}{\begin{list}{{\rm \arabic{muni}.}}
{\usecounter{muni}\setlength{\leftmargin}{0pt}
\setlength{\itemindent}{38pt}}}{\end{list}}

\newcommand{\nc}{\newcommand}                  \nc{\nf}{\infty}    
\nc{\dst}{\displaystyle}                       \nc{\nnb}{\nonumber} 
\nc{\beq}{\begin{equation}}                    \nc{\eeq}{\end{equation}} 
\nc{\beqa}{\begin{eqnarray}}                   \nc{\eeqa}{\end{eqnarray}}
\nc{\brm}{\begin{remunerate}}                  \nc{\erm}{\end{remunerate}}
\nc{\barr}{\begin{array}}                      \nc{\earr}{\end{array}}
             \nc{\mc}{\mathcal}

\nc{\bs}{\backslash}        \nc{\nl}{\newline}      \nc{\mb}{\mathbb}
\nc{\qq}{\quad\quad}        \nc{\ol}{\overline}     \nc{\pt}{\partial}
\nc{\dg}{\dagger}

\nc{\alf}{\alpha}    \nc{\be}{\beta}        \nc{\ga}{\gamma}    \nc{\ka}{\kappa}
\nc{\de}{\delta}     \nc{\eps}{\epsilon}    \nc{\vtht}{\vartheta}
\nc{\om}{\omega}     \nc{\vp}{\varphi}    \nc{\vsi}{\varsigma}
\nc{\vrho}{\varrho}  \nc{\tht}{\theta}      \nc{\la}{\lambda}

\nc{\Om}{\Omega}     \nc{\Ga}{\Gamma}      \nc{\De}{\Delta}

\nc{\Log}{{\rm Log }}           \nc{\tg}{{\rm tg }}
\nc{\sh}{{\rm sh }}             \nc{\ch}{{\rm ch }}

\nc{\tr}{{\rm tr }}

\nc{\LIM}{\mathop{\smash{\rm LIM}}}

\title{\bf The hydrogen atom in electric  and \\ magnetic fields : 
Pauli's 1926 article}
\author{\\Galliano VALENT
\thanks{\noindent Universit\'e Paris 7-Denis Diderot et Laboratoire de Physique Th\'eorique et des Hautes Energies,
 Unit\'e associ\'ee au CNRS UMR 7589, 
 2 Place Jussieu, 75251 Paris Cedex 05.}}
\date{ }

\begin{document}
\maketitle
\begin{abstract}
\noindent   The results obtained by Pauli in his 1926 article 
on the hydrogen atom made essential use of the  
quantum dynamical $so(4)$ symmetry of the bound states. Pauli used this symmetry to 
compute the perturbed energy levels of an hydrogen atom in a uniform electric field 
(Stark effect) and in uniform electric and magnetic fields. Although the 
Stark effect on hydrogen has been studied experimentally, Pauli's results 
in mixed fields have been studied only for Rydberg states 
of rubidium atoms in crossed fields and for lithium atoms in parallel fields.                
\end{abstract}

\vfill
To appear in the American Journal of Physics \hfill  March 2002

\newpage
\section{Introduction}
In 1926 there appeared nearly simultaneously two articles by Schr\"{o}dinger \cite{Sc}  
and Pauli \cite{Pa} that solved one and the same central problem of the newly 
discovered quantum mechanics: the computation of the hydrogen atom spectrum. 
In most textbooks on quantum mechanics the eigenvalue problem is solved using  
Schr\"{o}dinger's approach with spherical coordinates and simultaneously diagonalizing 
the set of operators $\,H_0,\,\vec{L}\,^2,\,L_z.$  The main benefit of this approach 
is that one can deal, in principle, with any central potential and that the 
eigenstates have definite parity, which is of practical importance in the study of 
the selection rules for radiative transitions. However, a simple problem like the 
Stark effect, is quite cumbersome to deal with perturbatively, except for the 
low lying states. 

An essential aspect of the Coulomb problem is that the Schr\"{o}dinger 
equation is also separable in parabolic coordinates \cite{ll}. This 
super-separability is related to the conservation of an additional vector 
$\,\vec{m},$ the so-called Runge-Lenz (RL) vector, which is specific to 
the Coulomb potential. Despite its name, the discovery of this conserved vector in 
Newtonian mechanics goes back to Hermann in the 18th century, as noted by 
Goldstein \cite{Go}, who mentions also early works by Laplace and Hamilton. 

Pauli succeeded in defining a quantum extension $\,\vec{M}\,$ of the 
RL vector which is an observable, that is, an hermitian operator, and 
commutes with the Hamiltonian $\,H_0.$ This success enabled him to 
calculate the spectrum of the hydrogen atom by an abstract approach. 
The quantum extension of the RL vector was used to obtain the dynamical 
$\,so(4)\,$ symmetry of the bound states. This symmetry, combined with 
the general theory of angular momentum allows to express the energies 
of the hydrogen atom \cite{fn}.

The main drawbacks of Pauli's approach are that it works only for a Coulomb 
potential and that the eigenstates have, in general, no definite parity. 
However, in the same article, Pauli gave two non-trivial applications of his ideas: 
the first order perturbation of the energy 
levels under a uniform electric field (Stark effect) and in  the more complicated 
case of uniform electric and magnetic fields. Guided by an analogous relation 
from classical mechanics, Pauli guessed the following crucial relation between 
quantum expectation values :
\beq\label{P}
\langle\,2H_0\,\vec{R}\rangle=\frac 32\,\langle\vec{M}\rangle,\eeq
where $\,\vec{R}\,$ is the position operator and the expectation 
value should be taken in any eigenstate of the Hamiltonian $\,H_0.$ 
It is natural to call Eq. (\ref{P}) Pauli's identity and we 
will later discuss its origin.

Textbooks that discuss more advanced topics related to the quantum RL vector 
(see Refs. \cite{Sch}-\cite{Ol}), do not treat these interesting applications. 
The main aim of this note is to popularize Pauli's results. In Sec. II we introduce 
the quantum RL vector and use it to derive the spectrum of the hydrogen 
atom \cite{fn1}. In Sec. III we present a new derivation of the crucial identity
\beq\label{BB}
2H_0\,\vec{R}=\frac 32\,\vec{M}+\frac 1{i\hbar}\,[H_0,\vec{T}],\eeq
which implies Eq. (\ref{P}). It is interesting to note that Eq. 
(\ref{BB}) was first derived by a clever calculation of commutators 
by Becker and Bleuler \cite{bb} fifty years after Pauli's article. Our derivation 
will make apparent that it is the natural quantum extension of a classical relation. 
In Sec. IV we consider the hydrogen atom in the presence of uniform 
electric and magnetic fields. We then use Pauli's elegant approach to 
obtain the first-order perturbed levels. We conclude with a short account of 
the experimental checks, some of which are quite recent.

\section{Background material}
To set our notation (we stick closely to those of \cite{Sch}), let us recall the basic properties of the quantum RL vector. As far as 
possible, we will use upper case letters for quantum operators. We 
will write the hamiltonian of the hydrogen atom
\beq\label{H}
H_0=\frac{\vec{P}^2}{2\mu}-\frac {\ka}R,
\qq\qq \ka=\frac{q^2}{4\pi\eps_0},\eeq
where $\mu$ is the reduced mass, $q$ is the proton charge and 
$\,\ka=q^2/4\pi\eps_0.$  Although going from the classical angular 
momentum $\vec{l}=\vec{r}\wedge\vec{p}$ to the operator  $\vec{L}=\vec{R}\wedge\vec{P}$ is not ambiguous 
and leads to an 
hermitian operator $\vec{L},$ the situation is somewhat more 
complicated for the RL vector. Classically its definition
\beq\label{RLcl}
\vec{m}=\frac 1{\mu}\,\vec{p}\wedge\vec{l}-\ka\,\frac{\vec{r}}{r}\eeq
shows that there are quantization ambiguities due to the lack of 
commutativity of the operators $\vec{P}$ and $\vec{L}.$ There are two 
possible (but non-hermitian) corresponding operators, 
$\vec{P}\wedge\vec{L}$ and 
$(\vec{P}\wedge\vec{L})^{\dg}=-\vec{L}\wedge\vec{P}.$ Pauli observed 
that the simplest choice for $\vec{M}$ so that it is hermitian is
\beq\label{bm1}
\vec{M}=\frac 1{2\mu}(\vec{P}\wedge\vec{L}
-\vec{L}\wedge\vec{P})-\ka\,\frac{\vec{R}}{R}.\eeq
From these definitions we can check that both $\vec{L}$ and $\vec{M}$ 
are conserved at the quantum level, that is, that they 
commute with $H_0.$

The classical Poisson brackets involving $\vec{l}$ and $\vec{m}$ 
generalize to commutators of $\vec{L}$ and $\vec{M}$:
\beq\label{bm2}\barr{c}
[L_i,L_j]=i\hbar\,\eps_{ijk}\,L_k,\\[4mm] [L_i,M_j]=i\hbar\,\eps_{ijk}\,M_k,\\[4mm]\dst  
[M_i,M_j]=\left(\frac{-2H_0}{\mu}\right)\cdot i\hbar\,\eps_{ijk}\,L_k,
\earr\eeq
Two further important relations should be noted:
\beq\label{bm3}\barr{c}
\vec{L}\cdot\vec{M}=\vec{M}\cdot\vec{L}=0,\\[4mm]\dst 
\vec{M}\,^2=\left(\frac{2H_0}{\mu}\right)(\vec{L}\,^2+\hbar^2)+\ka^2.
\earr\eeq
The checks of the conservation of the RL vector and of Eqs. (\ref{bm2})  
and (\ref{bm3}) are quite involved: a detailed calculation may be found  
in Ref. \cite{GM} (p. 462) and in Ref. \cite{Ol} (p. 265).

Let us restrict ourselves to the subspace of the Hilbert space spanned by the  
bound states $|\psi\rangle$ with energy $E<0$. $H_0^{-1}$ and $(-H_0)^{-1/2}$ can 
be defined by
\[H_0^{-1}\,|\psi\rangle=E^{-1}\,|\psi\rangle, \qq\qq 
(-H_0)^{-1/2}\,|\psi\rangle=(-E)^{-1/2}\,|\psi\rangle.\]
We also define
\beq\label{mprime}
\vec{M}\,'=\left(-\frac{\mu}{2H_0}\right)^{1/2}\,\vec{M}.\eeq
Because $\,\vec{L}\,$ and $\,\vec{M}$ commute with $H_0$, the algebra 
in Eq. (\ref{bm2}) becomes
\beq\label{bm4}\barr{c}
[L_i,L_j]=i\hbar\,\eps_{ijk}\,L_k,\\[4mm] [L_i,M'_j]=i\hbar\,\eps_{ijk}\,M'_k, \\[4mm]
[M'_i,M'_j]=\,i\hbar\,\eps_{ijk}\,L_k,\earr\eeq
which we recognize as an $so(4)$ algebra with generators $\,\vec{L},\ \vec{M}'$. For 
further use we define the generators
\beq\label{bm5}\barr{c}
\vec{I}=\frac 12(\vec{L}+\vec{M}\,'),\\[4mm]
\vec{K}=\frac 12(\vec{L}-\vec{M}\,').\earr\eeq
Their commutation rules are quite instructive :
\beq\label{bm6}\barr{c}
[I_i,I_j]=i\hbar\,\eps_{ijk}\,I_k,\\[4mm] [K_i,K_j]=i\hbar\,\eps_{ijk}\,K_k,\\[4mm]
[I_i,K_j]=0,\earr\eeq
because they show that the operators $\vec{I}$ generate an $so(3)_I$ Lie algebra, 
while the operators $\vec{K}$ generate an $so(3)_K$ Lie algebra. Furthermore 
these two algebras are completely independent (commuting), and their Casimir 
operators (i. e. operators commuting with the whole algebra) are 
$\vec{I}\,^2$ for $so(3)_I$ and $\vec{K}\,^2$ for $so(3)_K.$

The correspondence given in (\ref{bm5}) between $\,\vec{L},\ \vec{M}'\,$ and 
$\,\vec{I},\ \vec{K}\,$ gives just the change of basis that displays the 
isomorphism $\,so(4)\sim so(3)_I\oplus so(3)_K.$ 
From the general theory of quantum angular momentum \cite{Sch}, we can find a 
basis common to the operators $\,\dst\vec{I}\,^2,\ \vec{K}\,^2,\ I_z,\ K_z$ such that
\beq\label{new1}
\left\{\barr{c}
\vec{I}\,^2\,|i,m_i;k,m_k\rangle=i(i+1)\hbar^2\,|i,m_i;k,m_k\rangle 
\\[4mm] 
I_z|i,m_i;k,m_k\rangle=m_i\,\hbar\,|i,m_i;k,m_k\rangle\\[4mm]
\vec{K}\,^2\,|i,m_i;k,m_k\rangle=k(k+1)\hbar^2\,|i,m_i;k,m_k\rangle
\\[4mm] 
K_z|i,m_i;k,m_k\rangle=m_k\,\hbar\,|i,m_i;k,m_k\rangle,\earr\right.
\eeq
with $i,k=0,\frac 12,1,\ldots\  m_i\in[-i,\ldots, +i],$ and  $m_k\in[-k,\ldots, +k].$
The first relation in (\ref{bm3}) implies that the operators $\vec{I}$ and 
$\vec{K}$ are constrained by
\beq\label{bm7}\barr{c}
\vec{L}\cdot\vec{M}\,'=0\\[4mm]
 \vec{I}\,^2=\vec{K}^2\earr\eeq
while the second relation can be written as : 
\beq\label{bm8}
-\frac 12\mu\ka^2(H_0)^{-1}=2(\vec{I}\,^2+\vec{K}\,^2)+\hbar^2,\eeq
showing that the states $|i,m_i;i,m_k\rangle,$ are also hamiltonian eigenstates. At 
the level of the eigenvalues, the operator relation in Eq. (\ref{bm8}) gives
\beq\label{en}
-\frac 12\mu\ka^2\frac 1{E_n^{(0)}}=4i(i+1)\hbar^2+\hbar^2=
(2i+1)^2\hbar^2,\eeq
so that the inverse energies are given by the eigenvalues of the 
Casimir operators of $so(4).$ Up to the identification of the 
principal quantum number $n=2i+1,$ we obtain for the energy eigenvalues
\beq\label{en1}
E_n^{(0)}=-\frac{\mu\ka^2}{2\hbar^2n^2},\eeq
where $n=2i+1=1,2,\ldots$ The eigenstates $|i,m_i;i,m_k\rangle$ are 
such that $m_i\in[-i,\ldots+i]$ and $m_k\in[-i,\ldots+i].$ The 
degeneracy is easily seen to be $(2i+1)^2=n^2.$

These states are not eigenstates of $\vec{L}\,^2.$ Indeed the relation 
$\vec{L}=\vec{I}+\vec{K}$ shows that $l$ may take 
any value between 0 and $2i=n-1,$ in agreement with the Schr\" odinger approach. Note 
that the complete set of commuting observables diagonalized here are $H_0,\,I_z,$ 
and $K_z$ while in Schr\" odinger's approach one takes $H_0,\,\vec{L}\,^2,$ and 
$ L_z.$ Note also that the angular momentum $\vec{L}$ is axial, while $\vec{M}$ is 
a true vector, so that $\,\vec{I},\,\vec{K}\,$ have no definite parity, and hence 
their eigenstates cannot have any definite parity. This absence of a 
definite parity is also obvious from the fact that the states  
$|i,m_i;i,m_k\rangle$ have no definite value for the orbital angular momentum $l.$ 
Let us conclude that even for parity invariant potentials  the eigenstates need 
not be of definite parity as can be seen in the discussion of the simpler case 
of a particle in a box, given in Ref. \cite{bfv} (p.328).

\section{From classical to quantum Pauli's identity}
Equation (\ref{P}) has a classical content, because it 
involves mean values. Indeed, for any classical periodic trajectory,  
the mean value of a dynamical quantity $f(\vec{r})$, is defined by
\beq\label{cq1}
\langle f\rangle=\frac 1T\ \int_0^T\ f\,dt=\frac{\mu}{\la T}
\,\int_{-\pi}^{+\pi}\ f\ r^2\,d\phi,\eeq
where $T$ is the period and $\la=\mu r^2d\phi/dt$ is the angular 
momentum. For the Coulomb case, we have
\beq\label{cq2}\barr{c}
\dst\frac{r_0}{r}=1+e\,\cos\phi,\\[4mm]
\dst e=\sqrt{1+2\frac{{\cal E}\la^2}{\mu\ka^2}},\\[4mm]
\dst T=2\pi\frac{\mu r_0^2}{\la(1-e^2)^{3/2}},\earr
\eeq
where $r_0=\la^2/\mu\ka,$ ${\cal E}$ is the total energy and 
$e$ the eccentricity of the orbit. We can show that the classical RL vector, lying 
in the trajectory plane, is directed from the origin to the perihelion, which 
lies on the $x$-axis. Its components (see Ref. \cite{GM} (p. 460) for the details) 
are given by
\beq\label{l1}
m_x=\ka e,\qq\qq m_y=m_z=0.\eeq
TIf we take $f=\vec{r},$ we have
\beq\label{cq3}\barr{l}
\dst\langle x\rangle=\frac{\mu\,r_0^3}{\la T}\int_{-\pi}^{+\pi}\ 
\frac{\cos\phi}{(1+e\cos\phi)^3}\,d\phi=-\frac{3\pi \mu e 
r_0^3}{\la T(1-e^2)^{5/2}},\\[6mm]
\dst\langle y\rangle=\frac{\mu\,r_0^3}{\la T}\int_{-\pi}^{+\pi}\ 
\frac{\sin\phi}{(1+e\cos\phi)^3}\,d\phi=0,\qq\langle z\rangle=0.\earr\eeq
Some algebra leads to
\beq\label{l2}
2{\cal E}\,\langle x\rangle=\frac 32\,\ka e,\qq\qq
\langle y\rangle=\langle z\rangle=0.\eeq
If we compare Eqs. (\ref{l1}) and (\ref{l2}) and identify the total 
energy with the classical hamiltonian $h_0,$ we obtain Pauli's relation in 
classical mechanics
\beq\label{cp1}
2h_0\,\langle \vec{r}\rangle=\frac 32\,\vec{m}.\eeq

Equation (\ref{cp1}) is an equality between mean values of classical physical 
quantities, because  the RL vector is conserved. It follows that the difference 
between these two classical physical quantities must be a total derivative 
(whose mean-value necessarily vanishes). So we need to find some 
vector $\vec{\tau}$ such 
that
\beq\label{cp2}
2h_0\,\vec{r}=\frac 32\,\vec{m}-\frac d{dt}\vec{\tau}.\eeq
Since on dimensional grounds the sought vector has dimension of 
momentum $\times$(length)$^2$, it must have the structure
\beq\label{new3}
\vec{\tau}=a\,(\vec{r}\cdot\vec{p})\,\vec{r}+b\,\vec{r}\,^2\,\vec{p}.
\eeq
The coefficients $a$ and $b$, fixed by enforcing relation (\ref{cp2}), give for  
$\vec{\tau}$ the expression
\beq\label{new4}
\vec{\tau}=\frac 12\,(\vec{r}\cdot\vec{p})\,\vec{r}-\,\vec{r}\,^2\,\vec{p},\eeq
which can be written more suggestively, using a Poisson bracket as 
\beq\label{cp3}
2h_0\,\vec{r}=\frac 32\,\vec{m}+\{h_0,\vec{\tau}\},\qq\quad
\{f,g\}=\sum_s\frac{\pt f}{\pt x_s}\,\frac{\pt g}{\pt p_s}-
\frac{\pt f}{\pt p_s}\,\frac{\pt g}{\pt x_s}.\eeq

The above relations involve classical (commuting) quantities ; to 
generalize Eq. (\ref{cp3}) to the quantum level, we try the following quantum extension
\beq\label{cp4}
2H_0\,\vec{R}=\frac 32\,\vec{M}+\frac 1{i\hbar}\,[H_0,\vec{T}],\eeq
where the quantities involved are now operators and $\vec{T}$ is the unknown 
quantum extension of the classical quantity $\vec{\tau}.$ We point out that in the 
left hand side of Eq. (\ref{cp4}), the position of the hamiltonian operator with 
respect to $\vec{R}$ is  important.  In this case $\vec{T}$ can be written in terms 
of three vector operators
\beq\label{new5}
\vec{T}=a 
\,\vec{R}\,(\vec{R}\cdot\vec{P})+b\,(\vec{R}\cdot\vec{R})\,\vec{P}+
i\hbar\,c\,\vec{R}.\eeq
If we require Eq. (\ref{cp4}) to be satisfied, we obtain the result \cite{fn2}
\beq\label{vecT}
\vec{T}=\frac 12 \,\vec{R}\,(\vec{R}\cdot\vec{P})-(\vec{R}\cdot\vec{R})\,\vec{P}
+i\hbar\,\vec{R}.\eeq
Pauli's relation (\ref{P}) follows from Eq. (\ref{cp4}) when we take 
its expectation value between eigenstates of $H_0.$ We can write
\beq\label{new6}
\langle 2H_0\,\vec{R}\rangle=2\,E_n^{(0)}\,\langle \vec{R}\rangle=
\frac 32\,\langle \vec{M}\rangle.\eeq
For further use we will write Eq. (\ref{new6}) as
\beq\label{cp5}
\vec{R}\sim\frac 32\,(2\,E_n^{(0)})^{-1}\ \vec{M},\eeq
The symbol $\sim$ indicates that the equality holds only when expectation values 
between eigenstates (with energy $\,E_n^{(0)}\,$) of the unperturbed hamiltonian 
are calculated.

\section{Electric and magnetic fields}
In the approximation of an infinitely heavy nucleus, $\mu=m$ the 
electron mass. If we also neglect the diamagnetic 
terms, the perturbation due to the uniform electric field $\vec{E}$ and magnetic field $\vec{B}$ writes
\beq\label{new7}
H_1=q\,\vec{E}\cdot \vec{R}+\frac q{2m}\,\vec{B}\cdot\vec{L},\eeq
where $q$ is the proton charge. 
For the first order computation we need the matrix elements 
$\langle i,m'_i;i,m'_k|H_1|i,m_i;i,m_k\rangle$ for states of definie energy $E_n^{(0)}\,$ where $i=(n-1)/2$ has some fixed value. Its first piece,  
using Pauli's identity, in the form given by Eq. (\ref{cp5}), we have  
\beq\label{new8}
\vec{E}\cdot \vec{R}\sim\frac 32\,(2E_n^{(0)})^{-1}\,\vec{E}
\cdot \vec{M}=-(-2mE_n^{(0)})^{-1/2}\,\vec{E}\cdot \vec{M}\,'=-
\frac 32\,n\,\frac{a_0}{\hbar}\,\vec{E}\cdot\vec{M}\,',\qq \eeq
with $a_0=\hbar^2/\mu\ka.$ It is then sufficient to use
\beq\label{new9}
\vec{M}\,'=\vec{I}-\vec{K},\qq\qq \vec{L}=\vec{I}+\vec{K},\eeq
to obtain
\beq\label{pert1}
H_1\sim\left(-\frac 32\,n\,\vec{\cal E}+\vec{\cal B}\right)\cdot 
\frac{\vec{I}}{\hbar}+
\left(\frac 32\,n\,\vec{\cal E}+\vec{\cal B}\right)\cdot
\frac{\vec{K}}{\hbar},\eeq
with $\vec{\cal E}=qa_0\,\vec{E},\vec{\cal B}=
\mu_B\,\vec{B},$ and $\mu_B=q\hbar/2m.$
Notice that dimensional analysis shows that $\,\vec{\cal E}\,$ 
and $\,\vec{\cal B}\,$ have the same dimensions as energy.

Let us now define
\beq\label{Pfin2}
\vec{\nu}_{\pm}=\frac{\pm\frac 32\,n\,\vec{\cal E}+\vec{\cal B}}
{\left|\left|\pm\frac 32\,n\,\vec{\cal E}+\vec{\cal B}\right|\right|},\qq 
E_{\pm}=\left|\left|\pm\frac 32\,n\,\vec{\cal E}+\vec{\cal B}\right|\right|=
\sqrt{\frac 94 \,n^2\,\,\vec{\cal E}\,^2\pm 
3\,n\,\vec{\cal E}\cdot\vec{\cal B}+\vec{\cal B}\,^2},\eeq
so that the perturbation can be written as
\beq\label{pert}
H_1\sim \frac{E_-}{\hbar}\ (\vec{\nu}_-\,\cdot\vec{I})+
\frac{E_+}{\hbar}\ (\vec{\nu}_+\cdot\vec{K}).\eeq
This perturbation is made up of two completely independent (commuting) 
pieces which can be diagonalized separately. Indeed, one can check the 
relation \cite{fn3}
\beq\label{rot}
\vec{\nu}\cdot\vec{I}=e^{\frac i{\hbar}\tht\vec{\mu}\cdot\vec{I}}
\,I_z\,e^{-\frac i{\hbar}\tht\vec{\mu}\cdot\vec{I}},\eeq
with
\[\vec{\nu}=\sin\tht(\cos\phi\,\vec{i}+\sin\phi\,\vec{j})+
\cos\tht\,\vec{k},\qq\quad 
\vec{\mu}\equiv\frac{\vec{\nu}\wedge\vec{k}}{||\vec{\nu}\wedge\vec{k}||}
=\sin\phi\,\vec{i}-\cos\phi\,\vec{j},\]
which states that a rotation of angle $\tht$ around the axis $\vec{\mu}$ 
transforms the vector $\vec{\nu}$ into the unit vector $\vec{k}$ along the $z$ axis. It can 
be checked using the well-known identity
\[e^{iA}\,B\,e^{-iA}=B+i\,[A,B]+\frac{i^2}{2!}\,[A,[A,B]]+\cdots\]
and the $so(3)_I$ algebra for the generators $\vec{I}.$ It 
follows that the operator 
$\vec{\nu}\cdot\vec{I}$ has the same eigenvalues as $\,I_z$, i.e.
 $\, m_i\hbar.$ Similar results are also valid for $\vec{K}.$ 
If we adapt this result to the perturbation given by Eq. (\ref{pert}) 
we can write
\beq\label{rot1}
\left\{\barr{lll}
\dst\vec{\nu}_-\cdot\vec{I}=e^{\frac i{\hbar}\tht_-\vec{\mu}_-\cdot\vec{I}}
\,I_z\,e^{-\frac i{\hbar}\tht_-\vec{\mu}_-\cdot\vec{I}},\qq & 
 \cos\tht_-=\vec{\nu}_-\cdot\vec{k},\qq & \dst\vec{\mu}_-=
\frac{\vec{\nu}_-\wedge\vec{k}}{||\vec{\nu}_-\wedge\vec{k}||},\\[6mm]
\dst \vec{\nu}_+\cdot\vec{K}=e^{\frac i{\hbar}\tht_+\vec{\mu}_+\cdot\vec{K}}
\,K_z\,e^{-\frac i{\hbar}\tht_+\vec{\mu}_+\cdot\vec{K}},\qq & 
 \cos\tht_+=\vec{\nu}_+\cdot\vec{k},\qq & \dst\vec{\mu}_+=
\frac{\vec{\nu}_+\wedge\vec{k}}{||\vec{\nu}_+\wedge\vec{k}||},\earr\right.\eeq
and we obtain the first order correction to the energies 
\beq\label{Pfin}
\Delta\,E_n^{(1)}=m_i\cdot E_- +m_k\cdot E_+.\eeq
Recall that the quantum numbers $m_i$ and $m_k$ take all integer or 
half-integer values between $-(n-1)/2$ and $+(n-1)/2.$

\subsection{The first order Stark effect}
For vanishing magnetic field, we can take the electric 
field along the $z$ axis. From Eq. (\ref{pert1}) we obtain the first order perturbed energies 
\beq\label{S}
\Delta\,E_n^{(1)}=-\frac 32\,qa_0\,|\vec{E}|\,n(m_i-m_k).
\eeq
The levels are split into $2n-1$ sub-levels, each with residual degeneracy $n-|m|.$ 
Eq. (\ref{S}) is the celebrated first-order Stark effect formula, which was obtained 
in 1926 simultaneously by Waller \cite{Wa}, Wentzel \cite{We}, and Epstein \cite{Ep} 
by perturbation calculations in parabolic coordinates. A more handy reference is 
Ref. \cite{ll}, where perturbative results up to second order. The reader is urged 
to compare these extensive calculations (even in first order) with the 
elegance of Pauli's.

The experimental results from Ref. \cite{lkb} are given in figure 1 for 
the hydrogen atom. The low lying states exhibit a nearly linear field dependence 
up to a value of $10^{-5}\ $ au (atomic unit), but this value decreases with 
increasing $n.$ This linear regime accounts also for some crossings 
of the energy levels corresponding to different values of $n.$  The second order 
corrections are also sizeable in this region for some levels.  
For higher field values the situation is much more complicated since the 
states are broadened by tunnelling effects or even ionized, and  
cannot be accounted for by perturbation computations around the bound states! 
This complicated region lies to the right of the solid curve.

\subsection{First order Zeeman effect}
For vanishing electric field, we can take the magnetic field along the 
$z$ axis and we are back to the Zeeman energies
\beq\label{Z}
\Delta E_n^{(1)}=\mu_B\,|\vec{B}|\,m,\eeq
where $m=m_i+m_k.$
As for the Stark effect the energy levels are split up into $2n-1$ levels, 
each with residual degeneracy $n-|m|.$

\subsection{Parallel versus crossed fields}
Pauli's result simplifies for the special case of crossed  
electric and magnetic fields, i. e. such that $\vec{E}\cdot\vec{B}=0.$   
The perturbed energies become
\beq\label{CF}
\Delta\,E_n^{(1)}=E_{\perp}\cdot m,\quad\mbox{with}\quad  
E_{\perp}=\sqrt{\frac 94 \,n^2\,(qa_0\,\vec{E})^2+(\mu_B\,\vec{B})\,^2}.\eeq
In some sense the system is still in a Zeeman-like regime, with the same 
splitting pattern, but with a more complicated mixed field dependence appearing 
in the factor $E_{\perp}.$ Equation (\ref{CF}) has been 
checked for Rydberg states  of rubidium atoms (large principal quantum number 
$n=34$), which are essentially hydrogen-like \cite{pdb}. The experimental results, 
taken from Ref. \cite{pdb}, are reproduced 
in Fig. 2. The different lines correspond to different values of $m.$
Note how small the window is for the electric field 
(between $0$ and $20\ $V/cm) and for the magnetic field (between $0$ and 
$6\ 10^{-2}\ $T) to observe Pauli's quantization, but the results are in 
good agreement with (\ref{CF}) for weak fields. 

An experiment dealing with parallel electric and magnetic fields has 
been described in Ref. \cite{clp} for lithium atoms. These results are quite 
interesting,  since the symmetry breaking is completely different from the crossed 
case, with no left over degeneracy. A quite large set of 
references to theoretical as well as experimental work may be found in \cite{rfw}.

The second order perturbative computation was done quite recently 
\cite{So},\cite{msw}. The most interesting aspects cover now the non-perturbative 
transition to chaos ($B\gtrsim 10 T$) which is quite harder to deal with. The 
interested reader should consult Ref. \cite{msw}.

\vspace{1cm}
{\bf Acknowledgments :} We discovered Pauli's quantization in the 
very nice problem given in \cite{bd} (p. 57). For the hydrogen atom,  
the perturbative computation is worked out for the states with $n=2$ 
in polar coordinates. We are happy to thank A. 
Laverne and J. Letessier for kind help in dealing with the figures, O. Betbeder 
for a critical reading of the manuscript, D. Delande for the recent 
references on the subject and the Referees for useful suggestions.

\newpage

\noindent FIGURE CAPTIONS :

\begin{figure}[h]
\caption[]{Stark effect for the hydrogen atom. An atomic unit (au) 
corresponds to an electric field of $5,14\ 10^9\ $v/cm. 
From Ref. \cite{lkb}}
\end{figure}

\begin{figure}[h]
\caption[]{Rydberg states of rubidium atoms in crossed electric 
and magnetic fields. The electric field unit is $1\ $V/cm and the magnetic field unit $10^{-2}\ $T. From Ref. \cite{pdb}}
\end{figure}
\vfill

\newpage

$$  $$
\centerline{Figure 1}
\vspace*{2cm}
%%%%%%%%%%%%%%%%%%%%%%%%%%%%%%% Figure 1
\begin{figure}[h]
\vspace*{4cm}
\centerline{\epsfig{width=10cm,angle=-1,figure=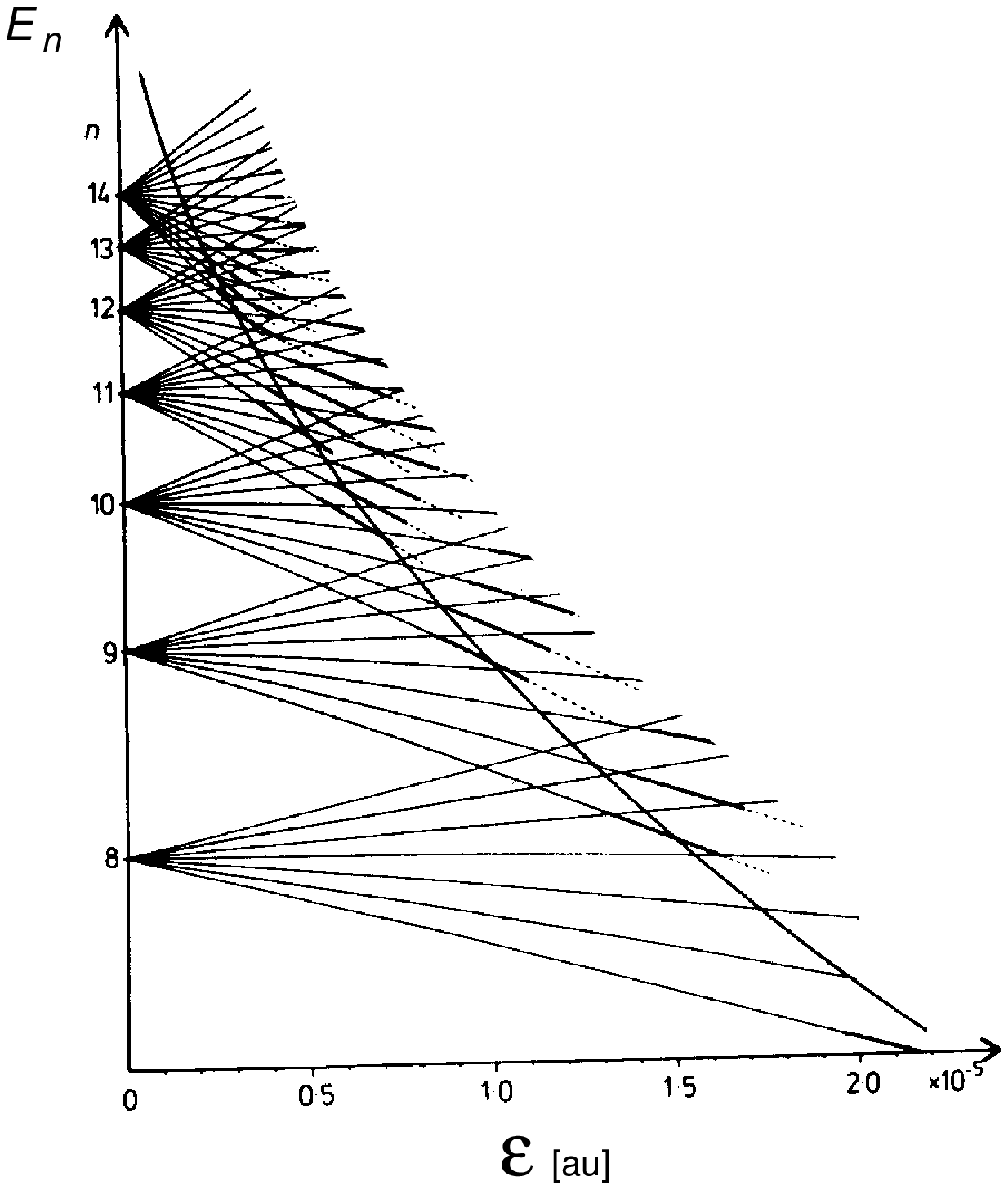}}
\end{figure}
%%%%%%%%%%%%%%%%%%%%%%%%%

\newpage

$$  $$
\centerline{Figure 2}
\vspace*{2cm}
%%%%%%%%%%%%%%%%%%%%%%%%%%%%%%% Figure 2
\begin{figure}[htb]
\vspace*{4cm}
\centerline{\epsfig{width=10cm,angle=-0,figure=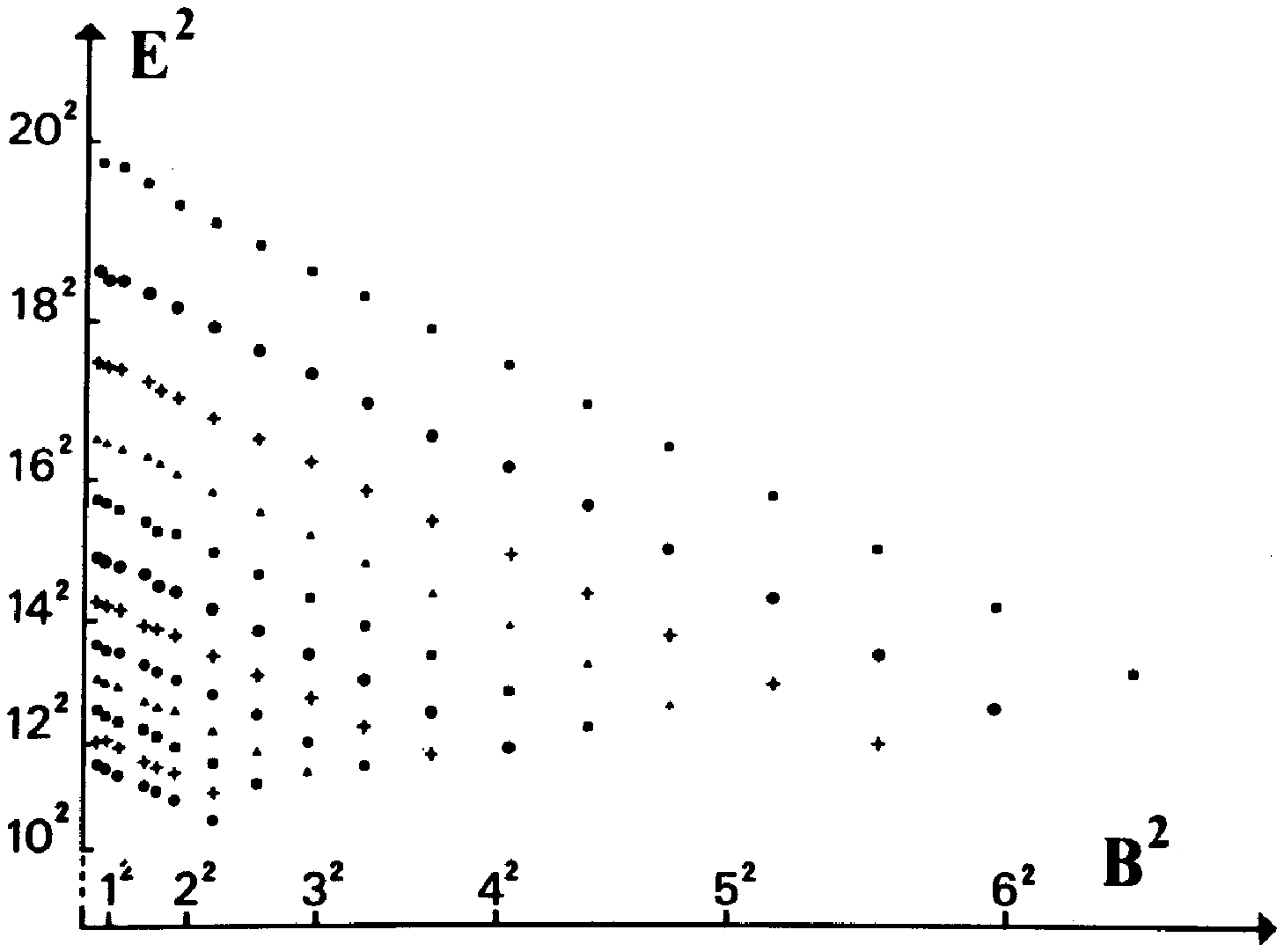}}
\vspace*{-0.cm} 
\end{figure}
%%%%%%%%%%%%%%%%%%%%%%%%%


\begin{thebibliography}{999}
\bibitem{fn}{As is well known, this analysis  
is equivalent to solving the Schr\"{o}dinger equation in parabolic coordinates.}
\bibitem{fn1}{This is reminiscent of the computation of the one dimensional 
harmonic oscillator spectrum using creation and annihilation operators instead of solving Schr\" odinger equation by power series.}
\bibitem{fn3}{We should not use this relation for $\tht=0,$ because in 
this case we have $\vec{\nu}\cdot\vec{I}=I_z$ and this operator is 
already diagonal. Similarly, for $\tht=\pi$ we have 
$\vec{\nu}\cdot\vec{I}=-I_z.$}
\end{thebibliography}
\end{document}